\documentclass[onecolumn,noshowpacs,floatfix,11pt,nofootinbib]{revtex4}
\usepackage{amssymb,amsmath}
\usepackage[dvips]{graphics,color}
\usepackage[colorlinks,hyperindex]{hyperref}
\usepackage{epsfig}\usepackage{float,stmaryrd,textcomp,bm,mathbbol}\usepackage{float,upgreek,yfonts}
\usepackage{graphicx,bigints,tikz}
\usepackage{mathrsfs}
\hypersetup{backref=true,pagebackref=true}
\hypersetup{%
  colorlinks = true,
  linkcolor  = blue,
  citecolor = cyan,
}
\newcommand{\M}{\mathcal{M}}
\usepackage{bookmark,textgreek}
\usepackage{hyperref,color,xcolor}
\hypersetup{hidelinks,hyperindex=true,colorlinks=true,breaklinks=true,urlcolor= blue}
\hypersetup{%
  colorlinks = true,
  linkcolor  = blue
}\usepackage{amssymb}
\usetikzlibrary{arrows,calc,shapes,decorations.pathreplacing}
\usepackage{caption}
\def\Cinf{\mathcal{C}^{\infty}}
\usepackage{capt-of}
\usepackage{faktor}
\def\Mat{\text{Mat}}
\usepackage{pdfpages}
\usepackage{url}
\def\cc{\mathbb{C}}

\def\hh{\mathbb{H}}

\usepackage[all]{xy}
\def\cl{\mathcal{C}\ell}
\def\Cl{\mathcal{C}\ell}
\def\cK{\mathrm{\cal K}}
\usepackage{hyperref}
\usepackage{xcolor}
\def\rr{\mathbb{R}}
\def\tcD{\mathring{\mathcal{D}}}
\newcommand{\beq}{\begin{eqnarray}}

\newcommand{\eeq}{\end{eqnarray}}
\begin{document}

\title{New generalized Killing spinor field classes in warped flux compactifications}

\author{D. Gonçalves Fabri}
\email{D.GoncalvesFabri1@universityofgalway.ie}
\affiliation{School of Mathematical and Statistical Sciences, University of
Galway, Galway H91 TK33, Ireland.}
\author{R. da Rocha}
\email{roldao.rocha@ufabc.edu.br}
\affiliation{Federal University of ABC, Center of Mathematics,  Santo Andr\'e, 09210-580, Brazil.}

\newcommand*\quot[2]{{^{\textstyle #1}\big/_{\textstyle #2}}}

\newcommand{\T}{\mathsf{T}}

\newcommand{\blkdiam}{\tikz\node[rectangle, draw, yscale=1, scale=.47, rotate=45, fill=black] {};}

\newcommand{\cliff}{{\mathcal{C}\ell}_{p, q}}

\newcommand{\tangb}{T^{*}\mathcal{M}}

\newcommand{\tangs}{T^{*}_x\mathcal{M}}

\newcommand{\extbund}{{\bigwedge}(T^{*}\mathcal{M})}

\newcommand{\extevbund}{{\bigwedge}^{+}(T^{*}\mathcal{M})}		

\newcommand{\extfib}{{\bigwedge}(T^{*}_x \mathcal{M})}		

\newcommand{\extevfib}{{\bigwedge}^+(T^{*}_x \mathcal{M})}	

\newcommand{\clifbund}{{\mathcal{C}\ell}(T^{*}\mathcal{M})}	

\newcommand{\clifevbund}{{\mathcal{C}\ell}^+(T^{*}\mathcal{M})}	

\newcommand{\cliffib}{{\mathcal{C}\ell}(T^{*}_x \mathcal{M})}	

\newcommand{\secextabrev}{\varGamma({\bigwedge})}

\newcommand{\secextkabrev}{\varGamma({\bigwedge}^k)}

\newcommand{\secext}{\varGamma({\bigwedge} (T^{*}\mathcal{M}))}

\newcommand{\secextk}{\varGamma({\bigwedge}^k (T^{*}\mathcal{M}))}

\newcommand{\unsecex}{1_{\varGamma}}

\newcommand{\secp}{\varGamma(\mathcal{P})}

\newcommand{\secpe}{\varGamma(\mathcal{P}^+)}

\newcommand{\secpm}{\varGamma(\mathcal{P}^\pm)}

\newcommand{\secs}{\varGamma(\mathcal{S})}

\thispagestyle{empty}

\begin{abstract}
The generalized Fierz identities are addressed in the K\"ahler--Atiyah bundle framework from the perspective of the equations governing constrained generalized Killing spinor fields. We explore the spin geometry in a Riemannian 8-manifold, $\M_8$, composing a warped flux compactification AdS$_3\times \M_8$, whose metric and fluxes preserve one supersymmetry in AdS$_3$. Supersymmetry conditions can be efficiently translated into spinor bilinear covariants, whose algebraic and differential  constraints yield 
identifying new spinor field classes. Intriguing implications and potential applications are discussed. \end{abstract}

\maketitle

\section{Introduction}
\label{sec:intro}

Clifford algebras and their classification provide an intimate relationship between division algebras and supersymmetry \cite{Bonora:2009ta}. 
The classical approach defines spinors as objects carrying irreducible representations of the classical Spin group, which is a restriction of the twisted Clifford--Lipschitz group to multivectors of unit norm in the associated Clifford algebra  \cite{oxford}. The Atiyah--Bott--Shapiro mod 8 classification of real 
 Clifford algebras induce classical real spinors to be also mod 8-classified \cite{Atiyah:1964zz}.
An ulterior relevant classification allocates classical spinors into disjoint classes when the spinor bilinear covariants are taken into account, satisfying the generalized Fierz identities for any finite-dimensional spacetime endowed with a metric of arbitrary signature \cite{bab1}. Spinor field classifications  in several  dimensions and metric signatures have been reported in the context of compactifications  underlying supergravity and string theory, as the more usual AdS$_5\times S^5$ and AdS$_4\times S^7$  compactifications, in Refs. \cite{Bon14,brito,Yanes:2018krn}. It implements  new recently obtained fermionic solutions in string theory and AdS/CFT \cite{Meert:2018qzk}.

These more general spinor field classifications,   according to the bilinear covariants, generalize Lounesto's spinor field classification in Minkowski spacetime, which, besides encompassing Dirac,  Majorana, and Weyl spinors, also encloses the Penrose flag-dipole, flagpole, and dipole spinor constructions.
Some of these spinors can be used to construct mass dimension one spinor fields, which 
have been reported to consistently account for the dark matter problem \cite{Bernardini:2012sc,Ahluwalia:2022yvk,deGracia:2023yit}. In Minkowski spacetime, spinors can be reconstituted from their bilinear covariants, as stated by the tomographic reconstruction theorem  \cite{Takahashi:1982bb,Crawford:1985qg,Mosna:2003am}, whose higher-dimensional generalization was addressed in Ref. \cite{Bonora:2015ppa}. The reconstruction theorem for real spinors from spinor bilinears is subject to global obstructions \cite{LazaroiuRec}.

We are here motivated to better explore supersymmetric AdS$_3$ backgrounds in $M$-theory, in appropriate compactifications of supergravity, where gauge/gravity holographic duality can be properly addressed. Strongly-coupled 2-dimensional conformal field theories (CFT$_2$) are dual to weakly-coupled gravity in AdS$_3$ backgrounds. CFT$_2$ offers more possibilities from the (super)algebraic point of view. The Kachru--Kallosh--Linde--Trivedi (KKLT) formalism was introduced on supersymmetric AdS$_4$ vacua, constituting a landscape in the context of string theory \cite{Kachru:2003aw}. The analogous landscape consisting of supersymmetric AdS$_3$ backgrounds, being a relevant problem to tackle, can also shed new light on the existing KKLT approach \cite{Ashmore:2022ydf}.
At low energies, the gauge/gravity correspondence establishes how gravitational dynamics in a bulk, dictated by Einstein's field equations, relate to fluid dynamics, governed by the Navier--Stokes relativistic equations on the boundary.
We have recently implemented a duality between incompressible viscous fluids and gravitational backgrounds with soft-hair excitations through suitable boundary conditions on gravitational backgrounds. It establishes a correspondence between generalized incompressible Navier--Stokes equations and black hole horizons with soft-hair \cite{Ferreira-Martins:2021cga}. These results are also based on the recent developments by  Hawking, Perry, and Strominger, who showed that non-extremal stationary black holes exhibit infinite-dimensional symmetries in the near-horizon region, known as supertranslations \cite{Hawking:2016msc}. This setup contributes to solving the information paradox for black holes. These symmetries are similar to the ones arising in asymptotically flat spacetimes at null infinity, known as Bondi--van der Burg--Metzner--Sachs (BMS)  symmetry \cite{Bondi:1962px,Sachs:1962zza}. 
The corresponding algebra is an infinite-dimensional extension of the Poincaré algebra. Recently the BMS algebra was reported in the extension of AdS/CFT to asymptotically flat spacetimes, playing a central role in the holographic description of black holes. 
The prototypical example is the microscopic derivation of the entropy of an asymptotically black hole  in AdS${}_3$ in terms of the
Virasoro algebra \cite{Brown:1986nw}, which also appears in the description of  warped AdS$_3$ black hole geometries \cite{Henneaux:2011hv}. 
Supertranslations produce conservation laws and require black holes to carry a large amount of
hair \cite{Grumiller:2019fmp}. 
Soft hair is implemented by smooth bosonic fields on the black hole horizon  and controls the final stages of the  evaporation of a black hole \cite{Mirbabayi:2016axw}, based on Weinberg's soft graviton theorem \cite{Weinberg:1965nx}. Some solutions in supergravity in eleven dimensions can attain an AdS$_3$/CFT$_2$ gauge/gravity dual description. The supersymmetric dual to the CFT$_2$  corresponds to a set of superconformal algebras, which is larger than the usual higher-dimensional supersymmetric CFT \cite{Grumiller:2016pqb}. The Virasoro algebra underlying CFT$_2$, being infinite-dimensional, leads to many possibilities for supersymmetric extensions \cite{Fradkin:1992bz,Legramandi:2020txf}. The classification of spinor fields in compactifications involving AdS$_3$ can bring relevant new information in gauge/gravity dualities, accordingly. Ref. \cite{Beck:2017wpm} approached the classes of superconformal algebras that may be embedded into the AdS$_3$ compactification component of solutions in ten and eleven-dimensional supergravity. The CFT$_2$ preserving maximal supersymmetry for AdS$_3$ solutions in  eleven dimensions was studied \cite{Haupt:2018gap}, also motivating the construction of a spinor field classification in the  compactified space complementary to AdS$_3$. Since 2-dimensional superconformal algebras are chiral, supersymmetric AdS$_3$ solutions can support two algebras of opposite chirality. Other relevant superconformal algebras were reported in this context in Refs. \cite{Ivanov:2015iia,Aizawa:2013rra,Kuznetsova:2011je,Aizawa:2018wxj,Berkovits:2008ic}.

\textcolor{black}{Extended theories of gravity are quantum gravity candidates and may be viewed as a setup to explore several topology and uniqueness theorems regarding classical solutions in gravity. Additionally, in this context, semiclassical thermodynamic aspects for computing black hole entropy and the area law can be considered. 
AdS$_3$ Einstein's gravity is a framework exploring  the interplay between CFT, gravity, and spacetime geometry, exhibiting unique features, including the existence of a topological phase. AdS$_3$ Einstein's gravity can be formulated by a dual two-dimensional CFT on its boundary, in the AdS/CFT setup. 
Some aspects of gravity in AdS$_3$ can be better  understood from the point of view of the boundary CFT. 
The three-dimensional gravity theory is renormalizable, permitting quantum effects to be explored. Solutions to AdS$_3$ Einstein's gravity with boundary conditions have a symplectic structure, and the Brown-Henneaux asymptotic symmetries become symplectic symmetries in AdS$_3$ \cite{Brown:1986nw}. 
The near-horizon decoupling limit preserves the chiral half of the symplectic symmetries. }

\textcolor{black}{Ref. \cite{Strominger:1997eq} addressed  
AdS$_3$ black holes whose near-horizon geometries are  AdS$_3$, and used the fact that quantum
gravity on AdS$_3$ is a CFT, to microscopically compute the black hole entropy, with a precise numerical agreement with the
Bekenstein--Hawking formula. It sheds new light on the information puzzle, as information is more safely stored on the black hole surface than in the black hole inner region, which is causally inaccessible. Statistical accounting for the
entropy sometimes suffers from ultraviolet divergences in quantum
gravity, added by the infinite number of low-energy modes. These issues can be resolved for black holes whose near-horizon geometry is locally AdS$_3$, including the three-dimensional BTZ black hole \cite{Banados:1992wn}. The black hole entropy can also be derived microscopically by summing up excitations of the AdS$_3$ spacetime.}

Consider an arbitrary  orientable manifold of finite dimension and arbitrary metric signature. The topological conditions for the existence of spinor bundles in indefinite signature are related to the modified second Stiefel-Whitney classes \cite{Karoubi}. Within this setup, the existence of spinor bilinear covariants relies upon real,  complex, or quaternionic structures that are compatible and inherent to the respective dimension and metric signature, but still have an underlying mod 8 Atiyah--Bott--Shapiro  periodicity with respect to real Clifford algebras.
In other words, depending on the 
dimension and metric signature, some homogeneous differential forms, playing the role of bilinear covariants, can vanish due to algebraic obstructions. Therefore, some of the  spinor bilinear covariants can attain null values. This property is dictated by generalized geometric Fierz identities. Despite the natural severe constraints in most dimensions and metric signatures, Lounesto's spinor field classification on four-dimensional Lorentzian manifolds \cite{lou2} can be thrivingly promoted to other  
dimensions and metric signatures, which have outstanding importance in the investigation of fermionic fields in flux compactifications.
Moufang loops on the 7-sphere  composing the compactification AdS$_4\times S^7$ were studied in Refs. \cite{Bon14,Yanes:2018krn}, where new spinor classes have been found. These spinor fields were shown to correctly transform under the Moufang loop generators on $S^7$. On the other hand, new spinor field classes in the  compactification AdS$_5\times S^5$ were derived and investigated in Ref. \cite{brito}, representing new fermionic solutions in the context of AdS/CFT.  Ref. \cite{Lopes:2018cvu} reported new classes of spinor fields in the cone and cylinder formalisms, addressing compactifications of $M$-theory with one supersymmetry.

The K\"ahler--Atiyah bundle plays a fundamental role in the essence of the  bundle of real spinors. It provides a robust framework with efficient techniques to analyze the  geometric Fierz identities arising from supersymmetry conditions for flux compactification backgrounds \cite{bab2}.  Within this setup, one can investigate the quantity of preserved supersymmetries in AdS$_3$. As long as the spinor fields adhere to the constrained generalized Killing (CGK) conditions, the space of spinorial solutions can be redefined in terms of algebraic and differential equations involving bilinear covariants. Therefore, spinor bilinear covariants can be constrained by the generalized Fierz identities, mirroring the methodology employed in Lounesto's classification, however, in the context of the AdS$_3\times \M_8$ compactification. Through an appropriate combination of the bilinear pairing on the bundle of real spinors, new classes of spinor fields can be identified and discussed.

\textcolor{black}{Our main goal in this work consists of providing 
the main tools for constructing the fermionic sector of fluid/gravity duality in the context of the AdS$_3\times \M_8$ compactification. 
We aim to pave the formal construct for it. 
Fermionic modes can be naturally regarded, with fermionic corrections to the fluid on the boundary \cite{Erdmenger:2013thg}. 
Refs. \cite{Gentile:2012jm,Gentile:2013gea} reported fermionic wigs, implementing counterparts of black hole hairs, considering the full AdS/CFT duality relating supergravity and
its dual theory, yielding supersymmetric extensions of Navier--Stokes equations. More precisely, Refs. \cite{Gentile:2011jt,Grassi:2011wt} generalized the fluid/gravity correspondence to supergravity, considering supersymmetric fluids on the boundary. 
In the classical bosonic sector of fluid/gravity correspondence, one starts with a solution of gravity, usually either a black hole or a black brane, which corresponds to a solution in supergravity with all fermionic zero
modes. Then isometries of AdS are implemented to yield a new
solution which depends upon some constant parameters that are promoted to fields on the AdS boundary. Although the solution no longer solves the equations of motion, novel PDEs emerge from this protocol, yielding the Navier--Stokes equations for the fluid on the boundary. AdS space is
equipped with superisometries, which introduce new constant parameters in the solution, which can be anew promoted to local fields on the AdS boundary.  Ref. \cite{Gentile:2013nha}  showed  that the solution no longer solves the supergravity equations, and new equations emerge from imposing them, but now regarding the complete supermultiplet, whose bosonic component is the black hole itself. In addition, the parameters regulating the superisometries must be promoted to local fields on the AdS 
boundary, which can be interpreted as quantities representing the fluid on the boundary.
In order to solve these problems, one can consider supergravity with
a cosmological constant term, where both the AdS$_3$ spacetime and the BTZ black hole are solutions. Ref. \cite{Gentile:2012tu} computed the entire  supermultiplet associated with the BTZ black
hole by performing a finite supersymmetric  transformation, implementing the computation of the complete orbit, corresponding to fermionic wigs, starting from the black hole
solutions. The most general form of fermionic wigs can be derived from transformations of Killing spinors in AdS$_3$ space. This procedure generates the complete supergroup of isometries of AdS$_3$ space. By promoting the fermionic parameters of the superisometries to local parameters on the boundary, and replacing the corresponding fields in the equations governing gravity, the Navier--Stokes equations are recovered, for the bosonic sector, and new PDEs are obtained for the fermionic degrees of freedom. On the other hand,  the solution can be replaced in the gravitino
equation, yielding a new set of Dirac-type equations for the fluid, regarding the fermionic degrees of freedom. Ref. \cite{Meert:2018qzk} consists of a successful approach considering new classes of spinor fields, emulating the flagpole and flag-dipole singular spinor fields in Lounesto's classification. They were introduced as new modes in the fermionic sector of fluid/gravity duality, emerging from the duality between the gravitino on the supergravity side and the phonino mode arising in supersymmetric hydrodynamics. We want to implement the formal aspects that can lead to such a construction for the AdS$_3 \times \M_8$ compactification.}

This paper is organized as follows:  Sec. \ref{sec2prep} is devoted to the fundamental setup, including a description of the Clifford bundle as the K\"ahler--Atiyah bundle.  In Sec.  \ref{sec3clif}, we describe the bundle of real spinors within this approach, and the geometric Fierz identities are constructed upon the definition of admissible pairings on the bundle of real spinors. We revisit Lounesto's spinor field classification, based on the bilinear covariants in Minkowski spacetime, together with a discussion on their main properties in Sec \ref{sec4bil}. It includes the fundamental concept of a Fierz aggregate and the reconstruction theorem in Minkowski spacetime.
Sec. \ref{sec5flux} delves into flux compactifications in supergravity on the AdS$_3 \times \M_8$ warped compactification,  focusing on the $\mathcal{N} =1$ supersymmetry with a single non-trivial spinor field solution in a Riemannian $8$-manifold, $\M_8$. In Sec. \ref{sec6new},  we emulate Lounesto's spinor field classification to the AdS$_3 \times \M_8$ compactification from the algebraic obstructions that force some of the $k$-form bilinear covariants to vanish. We reformulate the bilinear pairing to obtain $32$ new disjoint classes of spinor fields in $\M_8$. In Sec. \ref{sec7concl}, we conclude by presenting our final remarks and outlook.

\section{Preparations}\label{sec2prep}

Let $(\M,g)$ be an oriented pseudo-Riemannian manifold of dimension $n$. 
The Clifford bundle of differential forms on the pair $(\M,g)$ is denoted by $\mathcal{C}\ell(T^{*}\M) = \bigsqcup_{x \in \M} \mathcal{C}\ell(T^{*}_{x}\M, g^{*}_x)$, where the Clifford algebras on the cotangent bundle at $x\in\M$, $\mathcal{C}\ell(T^{*}_{x}\M, g^{*}_x)$, are also denoted by $\mathcal{C}\ell_{p,q}$, where  $(p,q)$ is the  signature of $g$. We identify the Clifford bundle $\mathcal{C}\ell(T^{*}\M)$ as the exterior bundle $\bigwedge T^{*}\M$ endowed with the geometric product  $\diamond: \bigwedge T^{*}\M \times \bigwedge T^{*}\M \to \bigwedge T^{*}\M$ whose induced action on sections $\Gamma(\M,\bigwedge T^{*}\M)$, which is again
denote by $\diamond$ for simplicity, satisfies the following relations for every $1$-form $v \in \Omega^{1}(\M)$ and $k$-form $\zeta \in \Omega^{k}(\M)$
\begin{equation}\label{Eq_cliffproduct}
      v \diamond \zeta = v \wedge \zeta + \sharp(v) \rfloor  \zeta,\hspace{0.5cm}         \zeta \diamond v = (-1)^{k} ( v \wedge \zeta - \sharp(v) \rfloor  \zeta),
\end{equation}
\noindent where $\rfloor$ is the left contraction such that $g(\zeta_1\rfloor\zeta_2,\zeta_3)=g(\zeta_1,\zeta_2\wedge\zeta_3)$, for all $\zeta_1, \zeta_2, \zeta_3\in \Omega(\M)$ and $\sharp(v)$ is the musical isomorphism $\sharp : \Gamma (\M, T^{*}\M) \to \Gamma (\M, T\M),$ with inverse $ \sharp^{-1}\equiv \flat: \Gamma (\M, T\M) \to \Gamma (\M, T^{*}\M)$, induced by the metric $g$,  raising and lowering  indexes, as $\sharp(v) = \sharp(v_{i}e^{i}) = g^{ij}v^{j}e_{j}.$ The bundle of algebras $(\bigwedge T^{*}\M, \diamond)$ is called Kähler--Atiyah bundle of $(\M,g) $ \cite{bab2}. The space $\Omega(\M)$ of all inhomogeneous smooth forms on $\M$, endowed with the geometric product $\diamond$, is an associative algebra with unity over the ring $\Cinf(\M,\rr)$, known as Kähler--Atiyah algebra of $(\M,g)$, and it satisfies the isomorphisms  $(\Omega(\M), \diamond) \simeq \Gamma(\M, \mathcal{C}\ell(T^{*}\M)) \simeq \Gamma(\M,\bigwedge T^{*}\M)$. K\"ahler introduced a new geometric product acting on exterior differential forms \cite{kahler}, which can endow the Grassmann algebra,  making it isomorphic to a Clifford algebra, as an associative algebra. \textcolor{black}{The K\"ahler product was deeply explored by Graf in  Ref. \cite{Graf} and the algebra equipped with these equivalent products has been called the K\"ahler-Atiyah algebra. The Clifford product can be emulated in this context through the contracted wedge product \cite{cep,Lopes:2017exe}. The fiber bundle formulation was paved in Ref. \cite{bab2}.  See also Ref. \cite{Fabri:2024sne}, for a detailed exposition of these approaches}.

Each of the intrinsic notions to Clifford algebras \cite{oxford} carries over to Clifford bundles. For instance, the even-odd decomposition of the $\mathbb{Z}_{2}$-graded algebra,
\begin{equation}
    \bigwedge T^{*}\M = \bigwedge T^{*}\M^{\textsc{even}} \oplus \bigwedge T^{*}\M^{\textsc{odd}},
\end{equation}
\noindent as well as the grade involution, reversion, and conjugation operators, respectively given for $\upalpha \in \Omega^{k}({\M})$ by
\begin{equation}\label{involution}
    \widehat{\upalpha} = (-1)^{k} \upalpha, \hspace{0.5cm}\qquad \widetilde{\upalpha} = (-1)^{\frac{k(k-1)}{2}} \upalpha, \hspace{0.5cm} \qquad \overline{\upalpha} = \widehat{\widetilde{\upalpha}} =  (-1)^{\frac{k(k+1)}{2}}\upalpha.
\end{equation}
The geometric product between forms of arbitrary degree is constructed by repeated application of Eq. \eqref{Eq_cliffproduct}. To express this product concisely, the contracted wedge product of order $d$ between two arbitrary forms $\upalpha, \upbeta \in \Omega(\M)$ introduced by Chevalley and Riesz \cite{Chevalley, Riesz} is defined inductively by 
\begin{align}
    \begin{aligned}
          \upalpha \wedge_{0} \upbeta &= \upalpha \wedge \upbeta,\\
     \upalpha \wedge_{1} \upbeta &= \sum_{i_{1}, j_{1} = 1}^{n} g^{i_{1}j_{1}} (e_{i_{1}}\rfloor \upalpha) \wedge_{0} (e_{j_{1}} \rfloor \upbeta),\\
     &\vdots\\
\upalpha \wedge_{d} \upbeta &= \sum_{i_{d}, j_{d} = 1}^{n} g^{i_{d}j_{d}} (e_{i_{d}}\rfloor \upalpha) \wedge_{d-1} (e_{j_{d}} \rfloor \upbeta).
    \end{aligned}
\end{align}
\noindent Consequently, for a $k$-form $\upalpha \in \Omega^{k}(\M)$ and a $l$-form $\upbeta \in \Omega^{l}(\M)$ with $k\leq l$, the geometric product $\diamond$ between $\upalpha$ and $\upbeta$ is defined as follows    
\begin{align}
    \begin{aligned}
          \upalpha \diamond \upbeta &= \sum^{k}_{d=0} \frac{(-1)^{d(k-d) + \left \llbracket  \frac{d}{2} \right\rrbracket}}{d!}  \upalpha \wedge_d \upbeta,\\
           \upbeta \diamond \upalpha &= (-1)^{kl} \sum^{k}_{d=0} \frac{(-1)^{d(k-d+1) + \left \llbracket  \frac{d}{2} \right\rrbracket}}{d!}  \upalpha \wedge_d \upbeta,
    \end{aligned}
\end{align}
\noindent where  $\left \llbracket  \frac{d}{2} \right\rrbracket$ represents the integer part of $\frac{d}{2}$ . In an orthonormal coframe $\{e^{1},\ldots, e^{n}\}$ the volume form is defined as being the $n$-form $\uptau_{n} = \text{vol}(\Omega(\M)) = e^{12\ldots n} = e^{1} \wedge e^{2} \wedge \cdots \wedge e^{n}$ and satisfies the following relation 
\begin{equation}
     \uptau_n \diamond \uptau_n = \begin{cases}
        +1, \text{ if } p - q \equiv_8 0,1,4,5 \\
        -1, \text{ if } p - q \equiv_8 2,3,6,7 
    \end{cases},
\end{equation}
\noindent where $(p,q)$, with $n =  p + q$, is the signature of the pseudo-Riemannian manifold $\M$. The volume form $\uptau_{n}$ is central if, and only if, $n$ is odd \cite{bab2}. Let one define  the operators $\uprho_{\pm} := \frac{1}{2} (1 \pm \uptau_{n}) \in  \Omega^{0}(\M) \oplus \Omega^{n}(\M).$ These algebraic objects  satisfy the following properties:
\begin{itemize}
    \item [\textit{1}.] $\uprho_{+} + \uprho_{-} = 1,$
     \item [\textit{2}.] $ \uprho_{\pm} \diamond  \uprho_{\pm} = \frac{1}{4} (1 \pm \uptau_{n})^{2} 
     = \begin{cases}
         \frac{1}{2} (1 \pm \uptau_{n}), \text{ if } p - q \equiv_8 0,1,4,5 \\
       \pm \frac{1}{2} \uptau_{n}, \text{ if } p - q \equiv_8 2,3,6,7 
    \end{cases},$ 
    \item [\textit{3}.] $\uprho_{\pm} \diamond \uprho_{\mp} = \frac{1}{4} (1 \pm \uptau_{n}) (1 \mp \uptau_{n})
    = \begin{cases}
         0, \text{ if } p - q \equiv_8 0,1,4,5\\
        \frac{1}{2} \uptau_{n}, \text{ if } p - q \equiv_8 2,3,6,7
    \end{cases}.$
\end{itemize}
\noindent Note that if $p - q \equiv_8 0,1,4,5$, then the $\uprho_{\pm}$ are two mutually orthogonal idempotents in $\Omega(\M)$. The Hodge duality operator is defined as being the  mapping $\star_n: \Omega^{k}(\M) \to \Omega^{n-k}(\M)$ such that $\star_{n}(\upbeta) = \tilde{\upbeta} \diamond \uptau_{n}$ for a $k$-form $\upbeta$. Two operators $P_{\pm}= \frac{1}{2} (1 \pm \star)$ can be defined through right $\diamond$-multiplication,  by setting for $\upalpha  \in \Omega(\M),\; P_{\pm} (\upalpha) = \frac{1}{2}( \upalpha \pm \star \upalpha)$. Furthermore, whenever $p - q \equiv_8 0,1,4,5$, the elements $P_{\pm}$ are complementary mutually orthogonal idempotents. 
\noindent The images $  \Omega(\M)_{\pm} := P_{\pm} \left(\Omega(\M)\right) = \Omega(\M)  \diamond \, \uprho_{\pm}$ give rise to the splitting $\Omega(\M)$ $=   \Omega(\M)_{+} \oplus  \Omega(\M)_{-}$.

\section{Clifford algebra approach to spinors}\label{sec3clif}

The bundle of real pinors $\mathcal{P}$ over the manifold  $\M$ can be defined as being the real bundle whose fibers are spaces that carry the irreducible representation of the fibers $\cl(T^{*}_{x}\M,g_{x}^{*})$ of the Clifford bundle $\cl(T^{*}\M)$, for all $x \in U \subset \M$, here $U$ denoting an open set in $\M$. The bundle of real pinors is equipped with a morphism $\upgamma: (\bigwedge T^{*}\M, \diamond) \to (\text{End}(\mathcal{P}), \circ)$ that maps the Kähler--Atiyah bundle of $(\M,g)$ to the bundle of endomorphisms of $\mathcal{P}$, where here $\circ$ represents the product in the space of  endomorphisms. The induced mapping on global sections, with the same notations, wits \cite{bab2}
\begin{equation}
    \upgamma : (\Omega (\M), \diamond) \to \Gamma\left(\M,\text{End}(\mathcal{P})\right), \circ).
\end{equation}
\noindent For each point $x \in \M$, the fiber $\upgamma_x$ is an irreducible representation of the Clifford algebra $(\bigwedge T^{*}_x\M, \diamond_x) \simeq \Cl_{p,q} $ in the 
$\mathbb{R}$-vector space $\mathcal{P}_x$, which denotes the fiber of $\mathcal{P}$ at the point $x \in \M$. A section of the bundle of real pinors is called a pinor field. Since we are interested in spinor fields, it is natural to consider the bundle of real spinors $\mathcal{S}$ of $(\M,g)$, consisting of a bundle of modules over the even Kähler--Atiyah bundle $(\bigwedge T^{*}\M^{\textsc{even}}, \diamond)$. The fibers of the bundle of real spinors comprise objects that carry the irreducible representation spaces of $\cl^{\textsc{even}}(T^{*}_x \M, g_{x}^{*})$ in $\cl^{\textsc{even}}(T^{*}\M)$, for all $x \in U$. Likewise, a spinor field is defined as a section of the bundle of real spinors and each fiber $\mathcal{S}_{x}$ arises from the mapping $\upgamma^{\textsc{even}}: (\bigwedge T^{*}\M^{\textsc{even}}, \diamond) \to (\text{End}(\mathcal{S}), \circ)$. The restriction $\upgamma^{\textsc{even}}$ of $\upgamma$ to the subbundle $\bigwedge T^{*}\M^{\textsc{even}} \subset \bigwedge T^{*}\M$ makes any bundle of real pinors $(\mathcal{P}, \upgamma)$ to drop into a bundle of real spinors $(\mathcal{S}, \upgamma^{\textsc{even}})$. Henceforth, for the sake of simplicity, the notation ($\mathcal{S}, \upgamma$) is employed to emphasize the approach to spinors.

Real Clifford algebras are classified based on the Atiyah--Bott--Shapiro mod 8 periodicity, as presented in the following Table \ref{table1} \cite{oxford}.
\begin{table}[H]
\begin{center}
\scalebox{1.1}{\begin{tabular}{c|c}
\hline \hline 
 $p - q \mod 8 $ & $\cl_{p,q}$                                                                                 \\ \hline\hline  \rule{0pt}{3ex}
$0$             & $\Mat(2^{\left \llbracket  \frac{n}{2} \right\rrbracket}, \rr)$                                                   \\\hline\rule{0pt}{3ex}
$1$             & $\Mat(2^{\left \llbracket  \frac{n}{2} \right\rrbracket}, \rr) \oplus \Mat(2^{\left \llbracket  \frac{n}{2} \right\rrbracket}, \rr)$     \\\hline\rule{0pt}{3ex}
$2$             & $\Mat(2^{\left \llbracket  \frac{n}{2} \right\rrbracket}, \rr)$                                                   \\\hline\rule{0pt}{3ex}
$3$             & $\Mat(2^{\left \llbracket  \frac{n}{2} \right\rrbracket}, \cc) $                                                  \\\hline\rule{0pt}{3ex}
$4$             & $\Mat(2^{\left \llbracket  \frac{n}{2} \right\rrbracket-1}, \hh)$                                                 \\\hline\rule{0pt}{3ex}
$5$             & \;$\Mat(2^{\left \llbracket  \frac{n}{2} \right\rrbracket-1}, \hh) \oplus \Mat(2^{\left \llbracket  \frac{n}{2} \right\rrbracket-1}, \hh)$ \\\hline \rule{0pt}{3ex}
$6$             & $\Mat(2^{\left \llbracket  \frac{n}{2} \right\rrbracket -1}, \hh) $                                               \\\hline \rule{0pt}{3ex}
$7$             & $\Mat(2^{\left \llbracket  \frac{n}{2} \right\rrbracket}, \cc)$\\\hline\hline 
\end{tabular}}
\caption{Real Clifford algebras classification. Hereon the notation $\Mat(r,\mathbb{K})$ accounts for the algebra of $r\times r$ matrices, whose entries are elements of the field $\mathbb{K}$.}
\label{table1}\end{center}
\end{table}
\noindent On the other hand, the classification of complex Clifford algebras $\cl_{\cc}(n)$ does not depend explicitly on the metric signature but only on the parity of the manifold dimension $n$, as shown in the Table \ref{table2}.
\begin{table}[H]
\begin{center}
\scalebox{1}{\begin{tabular}{c|c}\hline \hline 
$n = 2k$ & $\cl_{\cc}(2k) \simeq \Mat(2^{k}, \cc)$ \\ \hline 
$n = 2k+1$             &\; $\cl_{\cc}(2k+1) \simeq \Mat(2^{k}, \cc) \oplus \Mat(2^{k}, \cc)$ \\ \hline \hline     
\end{tabular}}  \caption{Complex Clifford algebras classification.}\label{table2}\end{center}
\end{table}

An arbitrary Clifford algebra consists of either a simple algebra or the direct sum of simple algebras. The largest group that can be defined in a Clifford algebra $\mathcal{C}\ell_{p,q}$ is the group $\mathcal{C}\ell_{p,q}^{*}$ of invertible elements. Notably, an important subgroup of $\mathcal{C}\ell_{p,q}^{*}$ is the twisted Clifford--Lipschitz group,  
\begin{equation}\label{eq_clgroup}
    \Sigma^{p,q} = \{ a \in \mathcal{C}\ell_{p,q}^{*} : \widehat{a}xa^{-1} \in \mathbb{R}^{p,q}, \; \text{for all} \;\; x \in  \mathbb{R}^{p,q}\}.
\end{equation}
The reduced Spin group is the group whose elements are the even elements of $\Sigma^{p,q}$ of unit norm \cite{oxford}, being the double covering of the special orthogonal group. A classical spinor is defined as an element that carries an irreducible representation space of the reduced Spin group.

Starting from the classification of Clifford algebras, irreducible representations of the even subalgebra $\mathcal{C}\ell_{p,q}^{\textsc{even}}$ can be immediately obtained by the well-known isomorphisms $\mathcal{C}\ell_{p,q}^{\textsc{even}}\simeq \mathcal{C}\ell_{q,p-1} \simeq \mathcal{C}\ell_{p,q-1}$ \cite{oxford}.  Hence, real and complex classical spinor fields can also be classified within this approach, as shown respectively in Tables \ref{table3} and \ref{table4} \cite{oxford}.
\begin{table}[H]
\centering
\scalebox{1.1}{
\begin{tabular}{c|c}
\hline\hline
$p - q \mod 8$ & $\text{Classical spinor space } \mathcal{S}_{p,q}$ \\
\hline \hline
$0$ & $\rr^{2^{\left \llbracket  \frac{n-1}{2} \right\rrbracket}} \oplus \rr^{2^{\left \llbracket  \frac{n-1}{2} \right\rrbracket}}$ \\\hline 
$1$ & $\rr^{2^{\left \llbracket  \frac{n-1}{2} \right\rrbracket}}$ \\ \hline 
$2$ & $\cc^{2^{\left \llbracket  \frac{n-1}{2} \right\rrbracket}}$ \\ \hline
$3$ & $\hh^{2^{\left \llbracket  \frac{n-1}{2} \right\rrbracket-1}}$ \\ \hline
$4$ & $\hh^{2^{\left \llbracket  \frac{n-1}{2} \right\rrbracket-1}} \oplus \hh^{2^{\left \llbracket  \frac{n-1}{2} \right\rrbracket-1}}$ \\ \hline
$5$ & $\hh^{2^{\left \llbracket  \frac{n-1}{2} \right\rrbracket-1}}$ \\ \hline
$6$ & $\cc^{2^{\left \llbracket  \frac{n-1}{2} \right\rrbracket}}$ \\ \hline
$7$ & $\rr^{2^{\left \llbracket  \frac{n-1}{2} \right\rrbracket}}$ \\ 
\hline \hline
\end{tabular}}\caption{Classical spinors classification: the real case.}\label{table3}\end{table}

\begin{table}[H]
\begin{center}
\scalebox{1}{\begin{tabular}{c|c} \hline\hline
$n = 2k$ & \;\;$\cc^{2^{k-1}} \oplus \cc^{2^{k-1}}$  \\ \hline 
$n = 2k+1$ \;\;            & $\cc^{2^{k}}$ \\\hline\hline       
\end{tabular}}  \caption{Classical spinors classification: the complex case. }\label{table4}
\end{center}
\end{table}

The mapping $\upgamma$ induced on sections is  defined on a local coframe by $\upgamma^{p} = \upgamma(e^{p}) \in \Gamma(U,\text{End}\,\mathcal{S})$ and  satisfies the morphism property   $\upgamma(\upalpha \diamond \upbeta) = \upgamma(\upalpha) \circ \upgamma(\upbeta)$ for all $\upalpha, \upbeta \in \Omega(\M)$. Moreover, 
\begin{equation}
    \upgamma(e^{m_1 \ldots m_k}) = \upgamma^{m_1 \ldots m_k} = \upgamma^{m_1} \circ \cdots \circ \upgamma^{m_k}. 
\end{equation}

Let $\alpha$ in $\Omega(\M)$ be an inhomogeneous differential form. One can represent it, with respect to a local coframe $\{e^{1},\ldots,e^{n}\}$, as follows
 \begin{equation}\label{app}
    \upalpha = \sum_{k=0}^{n} \upalpha^{k} =  \sum_{k=0}^{n} \frac{1}{k!} \upalpha^{k}_{m_1 \ldots m_k} e^{m_1 \ldots m_k},
 \end{equation}
 \noindent where $\upalpha^{k} \in \Omega^{k}(\M)$ and $\upalpha^{k}_{m_1 \ldots m_k}$ are $\mathcal{C}^{\infty}$-functions on $U \subset \M$. Applying $\gamma$ on Eq. (\ref{app}) yields
 \begin{equation}
    \upgamma (\upalpha) =\sum_{k=0}^{n} \frac{1}{k!} \upalpha^{k}_{m_1 \ldots m_k} \upgamma^{m_1\ldots m_k}.
\end{equation}
\noindent The surjectivity or the injectivity properties of the mapping $\upgamma$ are not always verified, being contingent on the classification of $\cl_{p,q}$ \cite{bab2}.

A non-degenerate bilinear mapping $B$ defined on the bundle of real spinors $(\mathcal{S},\upgamma)$ is said to be admissible if, for every $\upxi, \upxi' \in \Gamma(\M,\mathcal{S})$ and $\upalpha \in \Omega^{k}(\M)$, the following conditions hold \cite{Alekseevsky:2003vw, bab1}:
 \begin{itemize}
     \item [\textit{1.}]
         $B(\upxi, \upxi') = \upsigma(B) B(\upxi', \upxi)$
  such that $\upsigma(B) = \pm 1$ is the symmetry of $B$. The positive [negative] sign mimics a self-adjoint [anti-self-adjoint] bilinear mapping.
      \item [\textit{2.}] $B(\upgamma(\upalpha)\upxi, \upxi') = B(\upxi, (-1)^{\frac{(k(k-\upepsilon(B))}{2}}\upgamma({\upalpha}) \upxi')$, such that $\upepsilon(B) = \pm 1$ is the type of $B$. 
      \item [\textit{3.}] Whenever $p - q \equiv_8 0, 4, 6, 7$, the splitting spaces $\mathcal{S}^{\pm}$ of $\mathcal{S}$ with respect to $P_{\pm}$ are either: a) isotropic, where $B(\Gamma\mathcal{(M,S^{\pm})},\Gamma\mathcal{(M,S^{\pm})}) = 0$, or b) orthogonal, for which $B\left(\Gamma\mathcal{(M,S^{\pm})},\Gamma\mathcal{(M,S^{\mp})}\right) = 0$.
 \end{itemize}   
\noindent The above properties of $B$ depend on the dimension $n$ and the metric signature $(p,q)$ of the manifold. These relations can be found in Refs. \cite{bab1, Alekseevsky:2003vw}.

 The non-degenerated bilinear form $B$ induces a bundle isomorphism 
\begin{align}\begin{aligned}
 f \colon \Gamma(\M,\mathcal{S}) &\longrightarrow \Gamma(\M,\mathcal{S})^{*} \\
 \upxi &\longmapsto \begin{alignedat}[t]{2} 
& f(\upxi) & \colon \Gamma(\M,\mathcal{S}) & \longrightarrow \mathbb{R} \\
 & & \upxi' & \longmapsto B(\upxi',\upxi).
\end{alignedat}\end{aligned}
\end{align}
\noindent A natural bundle isomorphism is also given by 
\begin{align}\begin{aligned}
 h \colon \Gamma(\M,\mathcal{S}) \otimes \Gamma(\M,\mathcal{S})^{*}  &\longrightarrow \text{End}(\Gamma(\M,\mathcal{S})) \\
 \upxi \otimes T &\longmapsto \begin{alignedat}[t]{2} 
& h(\upxi  \otimes T) & \colon \Gamma(\M,\mathcal{S}) & \longrightarrow \Gamma(\M,\mathcal{S}) \\
 & & \upxi' & \longmapsto T(\upxi')\upxi.
\end{alignedat}\end{aligned}
\end{align}
\noindent The combination of those isomorphisms can define the following mapping:
\begin{equation}
    E := h \circ (1 \otimes f) : \Gamma(\M,\mathcal{S}) \otimes \Gamma(\M,\mathcal{S}) \longrightarrow  \text{End}(\Gamma(\M,\mathcal{S})),
\end{equation}
\noindent which induces the endomorphism, $ E_{\lambda_1 , \lambda_2} := E (\lambda_1 \otimes \lambda_2): \Gamma(\M,S) \longrightarrow \Gamma(\M,S)$, having the  explicit form as follows
\begin{align}
    \begin{aligned}
        E_{\lambda_1, \lambda_2} (\upxi) &= ((g \circ (1 \otimes f))(\lambda_1 \otimes \lambda_2))(\upxi)= (g(\lambda_1 \otimes f(\lambda_2)))(\upxi)\\
        &= (f(\lambda_2)\lambda_1)(\upxi)= B(\upxi, \lambda_2) \lambda_1.
    \end{aligned}
\end{align}
\noindent Moreover, for $\lambda_1,\lambda_2,\lambda_3,\lambda_4 \in \Gamma(\M,S)$ one has
\beq
       \!\!\!\!\!\!\! (E_{\lambda_1,\lambda_2} \circ E_{\lambda_3,\lambda_4})(\upxi) &=& E_{\lambda_1,\lambda_2}( E_{\lambda_3,\lambda_4}(\upxi)) 
        = E_{\lambda_1,\lambda_2} (B(\upxi, \lambda_4)\lambda_3)
        \\
        &=& B(\upxi, \lambda_4)E_{\lambda_1,\lambda_2}(\lambda_3)\nonumber= B(\upxi, \lambda_4) B(\lambda_3, \lambda_2) \lambda_1 \\
        &=&  B(\lambda_3, \lambda_2) B(\upxi, \lambda_4) \lambda_1
        = B(\lambda_3, \lambda_2) E_{\lambda_1, \lambda_4} (\upxi),
\eeq
\noindent which defines the generalized Fierz relation, as 
\begin{equation}\label{fie}
     E_{\lambda_1,\lambda_2} \circ E_{\lambda_3,\lambda_4} = B(\lambda_3,\lambda_2) E_{\lambda_1, \lambda_4}.
\end{equation}
Eq. (\ref{fie}) encodes the seed for constructing the non-trivial classes of spinor fields according to their bilinear covariants.
For the Riemannian manifold $\M_8$ to be approached in Sec. \ref{sec5flux}, $p - q \equiv_{8} = 0$; in this case, $\upgamma$ is bijective and one can take the inverse mapping $\upgamma^{-1}$ and define the Fierz isomorphism \cite{bab2}
\begin{equation}
    \check{E} = \upgamma^{-1} \circ E,
\end{equation}
\noindent which depends on the choice of admissible bilinear form $B$. For all $\upxi, \upxi' \in \Gamma(U,S)$ the explicit local expansion of $E$ is 
\begin{equation}\label{both}
    E_{\upxi, \upxi'} =  \frac{1}{2^{\left \llbracket  \frac{n+1}{2} \right\rrbracket}}\sum^{n}_{k=0} \frac{1}{k!} B(\upgamma_{m_k \ldots m_1}\upxi, \upxi')\upgamma^{m_1 \ldots m_k}.
\end{equation}
\noindent Applying $\upgamma^{-1}$ to both sides  of Eq. \eqref{both}, the local expansion for the Fierz isomorphism $\check{E}$ can be  expressed as
\begin{equation}
    \check{E} = \frac{1}{2^{\left \llbracket  \frac{n+1}{2} \right\rrbracket}} \sum^{n}_{k=0} \check{\mathbf{E}}^{(k)}_{\upxi, \upxi'},
\end{equation}
\noindent such that 
\begin{align}
    \begin{aligned}\label{eq563}
        \check{\mathbf{E}}^{(k)}_{\upxi, \upxi'} = \frac{1}{k!} B(\upgamma_{m_k \ldots m_1}\upxi, \upxi')e^{m_1 \ldots m_k} = \frac{1}{k!} \upepsilon(B)^{k} B(\upxi, \upgamma_{m_1 \ldots m_k}\upxi')e^{m_1 \ldots m_k},
    \end{aligned}
\end{align}
\noindent are the geometric Fierz identities.  From the symmetry properties of the admissible pairing $B$, specific constraints may arise on the geometric Fierz identities, forcing some of them to vanish. It, in turn, enables the classification of spinor fields based on those quantities.

\section{Bilinear covariants and spinor field classification in 4-dimensional spacetimes}
\label{sec4bil}
 Let $(\M,\eta)$ be a (oriented) manifold, with   tangent bundle $T\M$ and a metric $\eta: \Gamma(\M, T\M) \times \Gamma(\M, T\M) \to\mathbb{R}$, admitting an exterior bundle 
$\bigwedge T^{*}\M$ with sections $\Gamma(\M,\bigwedge T^{*}\M)$ endowed by the geometric product $\diamond$ (Eq. \eqref{Eq_cliffproduct}). 
Considering the 4-dimensional Minkowski spacetime $\M$, the set  $\{e^{\mu }\}$ hereon denotes a basis for the section of the coframe bundle
${P}_{\mathrm{SO}_{1,3}^{e}}(\M)$, constructed upon the component of the orthogonal group that is connected to the identity $e$. Classical \textcolor{black}{Dirac} spinor fields  carry the 
$\rho = {\left(\frac12,0\right)}\oplus {\left(0,\frac12\right)}$ representation of the component of the Lorentz group connected to the identity, $\mathrm{Spin}_{1,3}^{e}$. For arbitrary  
spinor fields  
$\uppsi \in \Gamma(\M, {P}_{\mathrm{Spin}_{1,3}^{e}}(\M))\times_{\rho}
\mathbb{C}^{4}$, denoting by $\star_4$ the Hodge duality operator, the bilinear covariants read
\begin{subequations}
\begin{eqnarray}
\sigma &=& \bar{\uppsi}\uppsi\,,\label{sigma}\\
\mathscr{J}_{\mu } e^{\mu }=\mathscr{J}&=&\left(\bar{\uppsi}\upgamma _{\mu }\uppsi\right)\, e
^{\mu}\,,\label{J}\\
\mathscr{S}_{\mu \nu } e^{\mu}\wedge e^{ \nu }=\mathscr{S}&=&\frac{i}{2}\left(\bar{\uppsi}[\upgamma_\mu, \upgamma_\nu]\uppsi\right) \, e^{\mu }\wedge  e^{\nu }\,,\label{S}\\
 \mathscr{K}_{\mu} e^{\mu }=\mathscr{K}&=&i\left(\bar{\uppsi}\left(\star_4\upgamma _{\mu }\right)\uppsi\right)
\, e^{\mu }\,,\label{K}\\\omega&=&-i\bar{\uppsi}(\star_4 \mathbf{1})\uppsi\,,  \label{fierz}
\end{eqnarray}\end{subequations}
where $\bar\uppsi=\uppsi^\dagger\upgamma_0$ is the Dirac-adjoint conjugation, $\upgamma
_{\mu
\nu }=\frac{i}{2}[\upgamma_\mu, \upgamma_\nu]$,  and the $\upgamma_{\mu }$ satisfies the Clifford algebra $\upgamma_{\mu }\upgamma _{\nu
}+\upgamma _{\nu }\upgamma_{\mu }=2\eta_{\mu \nu }\mathbf{1}$ of Minkowski spacetime. 
The scalar $\sigma$ and pseudoscalar $\omega$ bilinear covariants carry the $(0, 0)$ 
representation of the Lorentz group, whereas both the forms $\mathscr{J}$ and $\mathscr{K}$
carry the $\left(\frac{1}{2},\frac{1}{2}\right)$ representation of the Lorentz group. The 2-form $\mathscr{S}$ carry the $(1,0)\oplus(0,1)$ representation of the Lorentz group.   Solely focusing on Dirac's electron theory,  $\mathscr{J}$ is a spacetime-conserved current density arising from the U(1) symmetry due to Noether's theorem. Its temporal component, $\mathscr{J}_0 = \uppsi^\dagger\uppsi  = \|\uppsi\|^2 \geq 0$, wits the electron probability density, which does not equal zero for the electron in Dirac's theory.  The complete set of Fierz--Pauli--Kofink (FPK) identities read 
\begin{align}
\begin{aligned}\label{FIERZ}
\mathscr{K}\wedge\mathscr{J} &= (\omega - \sigma\star_4)\mathscr{S}, &\mathscr{J}^{2} &= \omega^{2} + \sigma^{2}, &\mathscr{K}^{2} + \mathscr{J}^{2} &= 0 = \mathscr{J}\cdot\mathscr{K}, \\
\mathscr{S}\lfloor\mathscr{J}  &= \omega\mathscr{K}, &\mathscr{S}\lfloor\mathscr{K} &= \omega\mathscr{J}, &(\star_4\mathscr{S})\lfloor\mathscr{J} &= -\sigma\mathscr{K}, \\
(\star_4\mathscr{S})\lfloor\mathscr{K} &= -\sigma\mathscr{J}, &\mathscr{S}\lfloor\mathscr{S} &= -\omega^{2} + \sigma^{2}, &(\star_4\mathscr{S})\lfloor\mathscr{S} &= -2i\omega\sigma(\star_4\mathbf{1}), \\
\mathscr{S}\mathscr{K} &= (\omega - \sigma\star_4)\mathscr{J}, &\mathscr{S}^{2} &= \left(\omega + i\star_4\sigma\right)^2.
\end{aligned}
\end{align} 

In Dirac's electron theory, the bilinear covariants have an unequivocal physical interpretation. Denoting by $q$ the electron charge, the term $q\mathscr{J}_0$ carries the interpretation of charge density, and the spatial components of the current density, $qc \mathscr{J}_k$,  are the spatial electrical current density. Besides, the spatial object $\frac{q\hbar}{2mc} \mathscr{S}^{ij}$ stands for the magnetic moment density,  aligning the torque on the electron from an external magnetic field. The mixed component $\frac{q\hbar}{2mc} \mathscr{S}^{0i}$ is the electrical moment density. The spacetime components $(\hbar/2) \mathscr{K}_\mu$ of $ \mathscr{K}$ are interpreted as the chiral current density in quantum field theory, which obeys a conservation law in the electron zero-mass limit. The interpretation of the scalar $\sigma$ and pseudoscalar $\omega$ bilinear covariants is less usual in the literature, except for the mass term $\sigma = \bar\uppsi\uppsi$ entering Dirac-like fermionic Lagrangians, which can also account for the electron self-interaction as well, proportional to the quadratic mass term. The FPK identity $\sigma^2 + \omega^2 = \mathscr{J}^2$ in  \eqref{FIERZ} is usually realized as a probability density for regular spinor fields \cite{oxford}. 
Due to the 4-vectorial nature of the bilinear covariant $\omega$, under parity $P$ and charge conjugation $C$, the pseudoscalar bilinear covariant can probe particle physics systems undergoing $CP$ violation.

Lounesto's classification consists of splitting the spinor fields according to bilinear covariants into six disjoint classes, wherein  $\mathscr{J}\neq 0$ in all six sets below, corresponding to a non-trivial spinor field $\uppsi$,  which are used to construct mass dimension 3/2 fermionic fields\footnote{The important case $\mathscr{J}= 0$, non-trivial spinor fields, was reported in Ref.  \cite{CoronadoVillalobos:2015mns} and accounts for mass dimension one spinor fields \cite{HoffdaSilva:2019ykt}.} \cite{lou2}:
\begin{subequations}\beq
1)&&\mathscr{S}\neq 0,\;\;\;\mathscr{K}\neq0,\;\;\;\sigma\neq0,\;\;\;\omega\neq0,\label{tipo1}\\
2)&& \mathscr{S}\neq 0,\;\;\;\mathscr{K}\neq0,\;\;\;\sigma\neq0,\;\;\;
\omega = 0,\label{tipo2}\\
3)&&\mathscr{S}\neq 0,\;\;\;\mathscr{K}\neq0,\;\;\;\sigma=0, \;\;\;\omega \neq0,\label{tipo3}\\
4)&&\mathscr{S}\neq 0,\;\;\;\mathscr{K}\neq0,\;\;\;\sigma=0,\;\;\;\omega=0,
\label{tipo4}\\
5)&&\mathscr{S}\neq0,\;\;\;
\mathscr{K}=0,\;\;\;\sigma=0,\;\;\;\omega=0,
\label{tipo5}\\
6)&&\mathscr{S}=0,
\;\;\; \mathscr{K}\neq0,\;\;\;\sigma=0,\;\;\;\omega=0.\label{tipo6}\eeq\end{subequations}
\noindent  
Classical singular spinor fields are objects in the subsets (\ref{tipo4}, \ref{tipo5}, \ref{tipo6}), which have, respectively, the Penrose's flag-dipoles, flagpoles, and
dipoles  structures.  
The 1-form fields $\mathscr{J}$ and $\mathscr{K}$, respectively given by (\ref{J}) and (\ref{K}),  play the role of two poles, and the 2-form bilinear covariant (\ref{S}), $\mathscr{S}$, regards a flag consisting of a 2-covector plaquette. Within this pictorial perspective, spinor fields in the set \eqref{tipo5} present a null pole, $\mathscr{K} = 0$, a non-vanishing pole structure, $\mathscr{J}\neq0$, and additionally a flag $\mathscr{S}\neq 0$. Therefore, they are named flagpole spinors. Spinor fields in the set (\ref{tipo4}) have two non-null pole structures instead, namely $\mathscr{J}\neq0$ and $\mathscr{K}\neq0$, as well as the flag structure, $\mathscr{S}\neq0$.  Hence, spinor fields in the set (\ref{tipo4}) have a flag-dipole structure and can be engendered by using the concept of  pure spinors \cite{oxford}. When spinor fields in the set \eqref{tipo6} are taken into account, their very definition establishes that $\mathscr{J}\neq0$ and  $\mathscr{K}\neq0$.  Nevertheless, in this case, the flag plaquette $\mathscr{S}$ equals zero. Therefore, with two poles and no flag structure, the set \eqref{tipo6} encompasses dipole structures. 
A reciprocal useful classification was engendered in Refs. \cite{Cavalcanti:2014wia,Fabbri:2016msm}, comprising the most explicit forms of spinor fields in each of Lounesto's classes.

This spinor field classification paved several proposals for non-standard fermionic in the literature. Flag-dipole spinor fields encode solutions of the Dirac equation in manifolds with Kalb--Ramond fields \cite{daRocha:2013qhu,Fabbri:2010pk,Vignolo:2011qt}.  Majorana and Elko uncharged fermions are representative spinor fields in the set \eqref{tipo5}, which can also allocate charged solutions of the Dirac equation in particular black hole backgrounds. 
Other types of flagpole spinor fields and new types of pole and flag spinor fields were addressed in Ref. \cite{CoronadoVillalobos:2015mns}. The classification of spinor fields according to their bilinear covariants has paved the way for new fermions in quantum field theory, including mass dimension one quantum fields consistently describing dark matter   \cite{Lee:2018ull,Rogerio:2022tsl,Ahluwalia:2022yvk,Ahluwalia:2022ttu,deGracia:2023yit,Fabbri:2010ws,Fabbri:2011mi,HoffdaSilva:2022mtq, fabbri}.

The $\mathbb{C}$-multivector field constructed upon the bilinear covariants,
\begin{equation}
\mathscr{Z}=2\left(\sigma+\mathscr{J}+i\mathscr{S}+i(\mathscr{K}+\omega)(\star_4\mathbf{1})\right),
\label{boom1}
\end{equation}
 is said to be a Fierz aggregate, 
when the homogeneous differential forms $\sigma, \omega, \mathscr{J},\mathscr{S}$, and $\mathscr{K}$ obey the FPK identities (\ref{FIERZ}).
Also, if the  Fierz aggregate is Dirac self-adjoint, namely, if it satisfies the condition  
\beq\label{boome}
\upgamma^{0}\mathscr{Z}^{\dagger}\upgamma^{0} =\mathscr{Z},\eeq  the Fierz aggregate is said to be a  boomerang, due to the $\upgamma^0$ operator \cite{lou2}. By taking 
a Weyl spinor, $\psi$, and constructing from it two Majorana spinors $\psi_\pm = \frac12(\psi+C(\psi))$, Penrose originally introduced flags as $\mathscr{Z}_\pm = 2(\mathscr{J}\mp i\mathscr{S})$ \cite{lou2,Penrose:1985bww}.

When either $\omega$ or $\sigma$ are non-vanishing, the spinor field is said to be a regular spinor. When both $\omega$ and $\sigma$ concomitantly equal zero,  the spinor field is singular \cite{BuenoRogerio:2020bdm,daRocha:2009gb} and the usual FPK identities (\ref{FIERZ}) are replaced by the more general expressions involving the Fierz aggregate, to wit
\begin{subequations}
\beq
\label{nilp}\mathscr{Z}^{2}  &=&\sigma \mathscr{Z},\\
\mathscr{Z}\upgamma_{\mu}\mathscr{Z}&=&\mathscr{J}_{\mu}\mathscr{Z},\\
\mathscr{Z}i\upgamma_{\mu\nu}\mathscr{Z}&=&\mathscr{S}_{\mu\nu}\mathscr{Z},\\
i\mathscr{Z}\left(\star_4\upgamma_{\mu}\right)\mathscr{Z}  &=&\mathscr{K}_{\mu}\mathscr{Z},\\
 -\mathscr{Z}\upgamma^{0}\left(\star_4\mathbf{1}\right)\mathscr{Z}&=&\omega \mathscr{Z}.\label{nilp1}
\eeq\end{subequations}
Spinor fields can be reconstructed from their associated 
bilinear covariants. The reconstruction theorem asserts that when an arbitrary non-trivial spinor field $\upxi$ is taken into account, such that 
$\upxi^{\dagger}\upgamma_{0}\uppsi\neq0$, the Fierz aggregate can be employed to reconstruct the original spinor $\uppsi$, up to a phase, as \begin{equation}
\uppsi=\frac{1}{\mathscr{N}}e^{-i\upalpha}\mathscr{Z}\upxi, \label{a1}%
\end{equation}
\noindent where $\mathscr{N}^2=\frac{1}{4}{\upxi^{\dagger}\upgamma^{0}\mathscr{Z}\upxi}$ and the U(1) phase reads 
$e^{-i\upalpha}=\frac{1}{\mathscr{N}}\upxi^{\dagger}\upgamma_{0}\uppsi$ \cite{Takahashi:1982bb,Crawford:1985qg,Mosna:2003am}. The inversion theorem was extended in Ref. \cite{Rogerio:2023kcp}. 
Relevant ramifications have been reported in Refs. \cite{Fabbri:2020elt,HoffdaSilva:2022ixq,Rogerio:2020trs,Cavalcanti:2020obq}. Several subclasses of regular spinors and most of the classes that constitute singular ones in the sets (\ref{tipo1}) -- (\ref{tipo6}) are not supported by the same physical  interpretation given to the electron in Dirac's theory. 

The vast physical possibilities of constructing a quantum field from the spinors in the sets (\ref{tipo1}) -- (\ref{tipo6}) have been explored. A second-quantized spinor field classification was reported in Ref. \cite{Bonora:2017oyb}. In any first-quantized quantum theory,  the reconstruction theorem makes one construct (up to a phase) spinor fields for each set, as long as the bilinear covariants are known \cite{Takahashi:1982bb,Crawford:1985qg}. Hence, the knowledge of the covariant bilinears -- or equivalently, of the Fierz aggregate -- is essentially correspondent
 to knowing the spinor field itself, up to a U(1) phase. When the second-quantization protocol sets in, new features arise, since the classification of quantum spinor fields according to their bilinear covariants highly depends on the Fierz aggregate and the $n$-point correlators  \cite{Bonora:2017oyb}.

\section{Flux compactifications in warped geometries}\label{sec5flux}

The (off-shell) underlying structure of $\mathcal{N} = 1$ supergravity in three dimensions has been long comprehended \cite{Howe:1977us,Brown:1979ma,Siegel:1979fr}. Plenty of relevant developments in minimal 3-dimensional supergravity, also 
including the $\mathcal{N} = 1$ massive case, were established  \cite{vanNieuwenhuizen:1985cx,Uematsu:1984zy,Howe:1995zm,Kuzenko:2011xg,Kuzenko:2016qwo}.
One can consider supergravity on an 11-dimensional manifold,  $\M_{11}$, endowed with a pseudo-Riemannian metric ${\mathring g}$.
The action governing supergravity accommodates a 3-form potential associated with
4-form field strength $\mathring{G}\in\Omega^4\left(\M_{11}\right)$ and the gravitino ${\Psi}_M$, described by a spin-$3/2$ real spinor field. An incipient approach to the gravitino in supergravity, using the quadratic spinor Lagrangian  and the spinor field classification was considered in Ref. \cite{daRocha:2009gb}.  When the bosonic sector of supersymmetric
 backgrounds is regarded, both the gravitino vacuum expectation value and its supersymmetry variation must equal zero. These conditions demand the existence of a non-trivial 
solution ${\mathring \upeta}$ of the first-order equation of motion,
\beq
\label{susy}
 \tcD_M {\mathring \upeta} = 0.
\eeq
Here ${\mathring \upeta}$ can be also thought of as being the supersymmetry generator consisting of a Majorana spinor field, carrying the irreducible representation of the Spin$_{1,10}$ group, seen as a smooth section of the spin
bundle ${\mathring S}$. Besides, capital letter 
 Latin indexes run in the range $0,\ldots,10$, and $\tcD_M$ denotes  the
supercovariant connection
\beq
\tcD_M{} = {{\nabla}}^{\mathring S}_M - \frac{1}{288} \left({\mathring G}_{PNRQ}
\mathring{\upgamma}^{PNRQ}{}_M- 8{\mathring G}_{MNRQ}\mathring{\upgamma}^{NRQ} \right),
\eeq
for the $\mathring{\upgamma}^M$ being the generators of $\cl_{1,10}$, with real Majorana unitary irreducible representation of 32 dimensions (see Table \ref{table1}, regarding the Clifford algebra classification), for which
${\mathring \upgamma}^{0\ldots10} = {\mathring \upgamma}^0\circ\cdots\circ {\mathring
\upgamma}^{1}$ plays the role of the volume element in $\M_{11}$,  and
\beq
{\mathbf{\nabla}}^{\mathring S}_M=\partial_M+\frac{1}{4}{\mathring {\Omega}}_{MNP}{\mathring \upgamma}^{NP}~~
\eeq
is the usual spin connection on ${\mathring S}$, induced by the standard 
Levi--Civita compatible connection of ($\M_{11}$, $\mathring{g}$). Ref. \cite{Martelli:2003ki} considered a compactification down to an
${\rm AdS}_3$ space, with cosmological running parameter $\Uplambda=-8\upkappa^2$, for 
$\upkappa\in\mathbb{R}_+$.  In this case, one can split $\M_{11}={\rm AdS}_3\times \M_8$, where  $\M_8$ is a Riemannian 8-manifold with defined orientation, equipped with a  metric $\mathtt{g}$. The warped metric on $\M_{11}$ 
 therefore reads
\beq\label{warp1}
\mathring{g} =\mathring{g}_{MN}dx^M dx^N = e^{2\Updelta}\left(ds^2_3+ \mathtt{g}_{mn} dx^m dx^n\right).
\eeq
The warp factor $\Updelta$ in Eq. (\ref{warp1}) is a $\mathcal{C}^\infty$-function on $\M$, whereas $ds_3^2$ denotes the AdS$_3$ metric. The ansatz for the 4-form field strength ${\mathring G}$ wits 
\beq\label{warp2}
\mathring{G} = e^{3\Updelta}\left(\uptau_3\wedge f_1+F_4\right),
\eeq
where $f_1=f_{m} e^m\in \Omega^1(\M_8)$, $F_4=\frac{1}{4!}F_{mnrs} e^{mnrs}\in \Omega^4(\M_8)$ and
$\uptau_3$ stands for the volume 3-form equipping ${\rm AdS}_3$. Lowercase Latin indexes label objects in $\M_8$ and run in the range 
$1$ to $8$. 
The equation of
motion and Bianchi identity for $\mathring{G}$ respectively read \cite{Ashmore:2022ydf}
\beq
&d\left(e^{3\Updelta} F_4\right)=0,\\ &e^{-6\Updelta}d\left(e^{6\Updelta}\star_8 f_1\right)-\frac12 F_4\wedge F_4=0,\\ &e^{-6\Updelta}d\left(e^{6\Updelta}\star_8 F_4\right)-f_1\wedge F_4=0,
\eeq where $\star_8$ denotes the Hodge star operator related to the 8-manifold $\M_8$ metric.  For the Majorana spinor field, ${\mathring \upeta}$, the following ansatz can be employed \cite{Babalic:2014fua},
\beq
{\mathring \upeta}=e^{\frac{\Updelta}{2}}\upeta, 
\eeq
with $\upeta=\psi\otimes \upxi$, for $\upxi$ standing for a real Majorana--Weyl  spinor on $\left(\M_8, \mathtt{g}\right)$, carrying the irreducible representation of Spin$_{8,0}$ \cite{bab1}, and $\psi$ denoting a Majorana spinor on the AdS$_3$, carrying the irreducible representation of Spin$_{1,2}$. Formally, $\upxi$ is an element in a section of the bundle of real spinors of $\M_8$, which by Table \ref{table3} is a real vector bundle of rank $16$ on $\M_8$. It also carries a representation of the Clifford algebra
$\cl_{8,0}$. As $p-q\equiv_8 0$ for $p=8$ and $q=0$, the normal simple case is regarded and the structure $\upgamma:(\bigwedge T^\ast
\M,\diamond)\rightarrow ({\rm End}(\mathcal{S}),\circ)$,   underlying the K\"ahler--Atiyah bundle, is an isomorphism. Ref. \cite{bab2} utilized the notation $\upgamma^p=\upgamma(e^p)$, for any local frame of $\{e^p\}$ on the cotangent bundle of $\M$. In the Euclidean $(8,0)$  signature,  there exists a Spin$(8)$-invariant admissible bilinear pairing $B$ on the bundle of real spinors $\mathcal{S}$, with $\upsigma(B)=1$ and $\upepsilon(B)=1$, defined in Section \ref{sec3clif},  
which plays the role of a scalar product. Now, given $\psi$ a Killing spinor on the ${\rm AdS}_3$ space, the
supersymmetry condition \eqref{susy} splits into constrained generalized Killing (CGK) conditions for the Majorana spinor field $\upxi$,
\beq
\label{par_eq}
\mathcal{D}_m\upxi = 0,\qquad\qquad \mathcal{Q}\upxi = 0,
\eeq
where $\mathcal{D}_m$ is a linear connection on $\mathcal{S}$ and $\mathcal{Q}\in \Gamma(\M, {\rm End}(\mathcal{S}))$ is
an endomorphism in the bundle of real spinors. Analogously to Refs. \cite{Martelli:2003ki,Tsimpis,Ashmore:2022ydf}, the Majorana spinor field $\upxi$ is not assumed to have definite
chirality.  The space of solutions of Eqs. \eqref{par_eq} is a
finite-dimensional $\mathbb{R}$-linear subspace $\cK(D,\mathcal{Q})\subset\Gamma(\M,S)$ of
smooth sections of the bundle of real spinors. Refs. \cite{bab2,Babalic:2014fua} focused on obtaining a set of metrics and fluxes on $\M_8$ preserving a fixed number of supersymmetries in AdS$_3$. Equivalently, the set of metrics and fluxes on $\M_8$ is consistent with the $s$-dimensional subspace $\cK(D,\mathcal{Q})$, for a given  $s\in\mathbb{N}$. The case of supergravity regarding ${\cal N}=1$ supersymmetry on 3-dimensional manifolds  was considered in Refs. 
\cite{Martelli:2003ki,Tsimpis,bab2,Babalic:2014fua,Becker:2003wb}, which reported   the
explicit expressions for $\mathcal{D}$ and $\mathcal{Q}$ in Eqs. \eqref{par_eq}, as
\beq
\label{susy1}
\mathcal{D}_m &=&{\mathbf{\nabla}}^S_m+ \frac{1}{4}f_p\upgamma_{m}(\star_8{\upgamma}^{p})+\frac{1}{24}F_{m n p q }\upgamma^{ n p q}+\upkappa (\star_8\upgamma_m),\\
\label{susy2}
 \mathcal{Q}&=&\frac{1}{2}\upgamma^m\partial_m\Updelta -\frac{1}{288}F_{m n p q}\upgamma^{m n p q}
-\frac{1}{6}f_p (\star_8\upgamma^p)
-\upkappa\upgamma^{{1\ldots8}},
\eeq
which are consistent with the compactification ans\"atze (\ref{warp1}, \ref{warp2}),  
where $\upgamma^{{1\ldots8}}=\upgamma^1\circ\ldots \circ\upgamma^8$. It is worth emphasizing that the last terms on the r.h.s of Eqs. (\ref{susy1},  \ref{susy2}) depend upon the ${\rm AdS}_3$ cosmological running parameter. 

Within this structure, one can explore the number of supersymmetries preserved in AdS$_3$, encoded in the dimension $s$ of $\cK(D,\mathcal{Q})$.
The space of solutions of Eqs. \eqref{par_eq} can be reframed in terms of equations involving the bilinear covariants 
${\boldsymbol{\check{E}}}^{(k)}_{\upxi,\upxi'}=\frac{1}{k!}B(\upxi,\upgamma_{m_1\ldots m_k}\upxi') e^{m_1\ldots m_k}$, as long as the  spinor fields $\upxi,\upxi'$ obey Eqs. 
\eqref{par_eq}. The spinor bilinear covariants can be constrained by generalized Fierz identities, emulating the case of Lounesto's classification to the AdS$_3\times \M_8$ compactification. From 
an appropriate combination of the bilinear form, we can obtain 32 new classes of spinor fields.   
The equations for the bilinear covariants can be obtained when  the algebraic constraints
$\mathcal{Q}\upxi=\mathcal{Q}\upxi'=0$ are rewritten as
\beq
B\left(\upxi,\left(\mathcal{Q}^\intercal\circ\upgamma_{m_1\ldots m_k}\pm \upgamma_{m_1\ldots m_k}\circ \mathcal{Q}\right) \upxi'\right) =0.
\eeq
Besides, the remaining constraints $D_m \upxi=D_m\upxi'=0$ are solved by an algorithm well posed in Ref.  \cite{Martelli:2003ki}. The most straightforward case $s=1$ for
${\cal N}=1$ supersymmetry in AdS$_3$ can be therefore approached, by demanding that Eqs. \eqref{par_eq} admit one non-trivial solution $\upxi$ \cite{bab2,Babalic:2014fua}.
One can hence consider CGK spinor equations \eqref{par_eq} on $(\M_8, \mathtt{g})$, assuming a 1-dimensional space of solutions, corresponding to $s=1$. For this case, the mod 8 equivalence $p-q\equiv_8 0$ yields the normal simple case.

\section{New classes of spinor fields}\label{sec6new}

We aim to emulate Lounesto's spinor field classification, briefly  discussed in Sec. \ref{sec4bil}, to the AdS$_3 \times \M_8$ compactification, addressed in Sec. \ref{sec5flux}. As paved in Sec. \ref{sec4bil}, 
algebraic obstructions can force some of the $k$-form bilinear covariants to vanish, depending on the manifold dimension and the signature of the metric that endows it.
The Fierz aggregate in $(\M_{8}, \mathtt{g})$ is expressed as the following inhomogeneous differential form
\begin{equation}
    \check{E} = \frac{1}{16} \sum_{k=0}^{8} \check{\mathbf{E}}^{(k)}. 
\end{equation}

\noindent  Considering the admissible pairing $B$ with $\upsigma(B)= \upepsilon(B) = +1$. Eq. \eqref{eq563} gives
\begin{equation}
    \check{\mathbf{E}}^{(k)}_{\upxi, \upxi'} = \frac{1}{k!}  B(\upxi, \upgamma_{m_1 \ldots m_k}\upxi)e^{m_1 \ldots m_k}
\end{equation}

\noindent for all $m_1, \ldots, m_k = 1,\ldots,8$ and  $k = 0,\ldots,8$. The cases $k=2,3,6,7$ yield  $\check{\mathbf{E}}^{(k)} = 0$. Indeed, 
\begin{align}
    \begin{aligned}
        B(\upxi, \upgamma_{m_1 \ldots m_k} \upxi) &=  \upsigma(B) B(\upgamma_{m_1 \ldots m_k} \upxi,  \upxi) =  \upsigma(B) (-1)^{\frac{k(k-\upepsilon(B))}{2}} B( \upxi, \upgamma_{m_1 \ldots m_k}\upxi).       
    \end{aligned}
\end{align}
Therefore, since $\upepsilon(B) = \upsigma(B) = +1$ for  $k = 2,3,6,7$  one has $B(\upxi, \upgamma_{m_1 \ldots m_k} \upxi) = 0$, and consequently $\check{\mathbf{E}}^{(k)} = 0$. On the other hand, for $k = 0,1,4,5,8$, the non-vanishing bilinear covariants 
\begin{subequations}
\beq
\label{nonvanishingbilinears}
        &&\check{\mathbf{E}}^{(0)} = B(\upxi, \upxi),\\
        &&\check{\mathbf{E}}^{(1)} = B(\upxi, \upgamma_{m}\upxi)e^{m},\label{nonvanishingbilinears1}\\
        &&\check{\mathbf{E}}^{(4)} = \frac{1}{4!} B(\upxi, \upgamma_{{m_1 \ldots m_4}}\upxi)e^{{m_1 \ldots m_4}},\\
        &&\check{\mathbf{E}}^{(5)} = \frac{1}{5!} B(\upxi, \upgamma_{{m_1 \ldots m_5}}\upxi)e^{{m_1 \ldots m_5}},\\
        &&\check{\mathbf{E}}^{(8)} = \frac{1}{8!} B(\upxi, \upgamma_{{m_1 \ldots m_8}}\upxi)e^{{m_1 \ldots m_8}},\label{nonvanishingbilinears2}
\eeq\end{subequations}
\noindent define a unique class of spinor fields \cite{bab2}. Namely, the set where the associated bilinear covariants are identified to the homogeneous $k$-forms, by $\check{\mathbf{E}}^{(k)} = 0$, for $k=2,3,6,7$, and $\check{\mathbf{E}}^{(k)} \neq 0$, for $k=0,1,4,5,8$. Incorporating the protocol in Refs. \cite{Bonora:2015ppa, bab1}, a complexification procedure of bilinear covariants can be studied in $\cl_{8,0}$. 
Let us consider the complex structure $J$ on the bundle of real spinors, with $J^2 = -I$ \cite{bab1, Alekseevsky:2003vw}. From the $\mathcal{S} = \mathcal{S}^{+} \oplus \mathcal{S}^{-}$ splitting, with $\mathcal{S}^{\pm} = P_{\pm}(\mathcal{S})$, one has $\mathcal{S} = \mathcal{S}^{+} \oplus J(\mathcal{S}^{+})$ since $J(\mathcal{S}^{\pm}) = \mathcal{S}^{\mp}$. Then, considering the spinor field $\upxi$ correspondingly written as  $\upxi_{R} + J(\upxi_{I})$, the bilinear pairing yields
\begin{align}\label{bc}
        B(\upxi, \upgamma_{{m_1 \ldots m_k}} \upxi) &= B\left(\upxi_{R} + J(\upxi_{I}), \upgamma_{{m_1 \ldots m_k}}\upxi_{R} + J(\upxi_{I})\right)\nonumber \\\nonumber
&= B(\upxi_{R}, \upgamma_{{m_1 \ldots m_k}} \upxi_{R}) 
+ B(\upxi_{R}, \upgamma_{{m_1 \ldots m_k}} J(\upxi_{I}))\\\nonumber &\qquad+  B(J(\upxi_{I}), \upgamma_{{m_1 \ldots m_k}} \upxi_{R})
+ B(J(\upxi_{I}), \upgamma_{{m_1 \ldots m_k}} J(\upxi_{I}))\\\nonumber
&=  B(\upxi_{R}, \upgamma_{{m_1 \ldots m_k}} \upxi_{R}) 
+ B(\upxi_{R},(J \circ \upgamma_{{m_1 \ldots m_k}} )\upxi_{I})\\\nonumber&\qquad+(-1)^{\frac{n(n+1)}{2}} B(\upxi_{I}, (J\circ\upgamma_{{m_1 \ldots m_k}}) \upxi_{R})
- (-1)^{\frac{n(n+1)}{2}} B(\upxi_{I}, \upgamma_{{m_1 \ldots m_k}} \upxi_{I})\\\nonumber
&=  B(\upxi_{R}, \upgamma_{{m_1 \ldots m_k}} \upxi_{R}) 
-  B(\upxi_{I}, \upgamma_{{m_1 \ldots m_k}} \upxi_{I}) \\ &\qquad +B(\upxi_{R},(J \circ \upgamma_{{m_1 \ldots m_k}} )\upxi_{I})  + B(\upxi_{I}, (J\circ\upgamma_{{m_1 \ldots m_k}}) \upxi_{R}),
\end{align}
\noindent where $J^{\intercal} = (-1)^{\frac{n(n+1)}{2}} J$ \cite{bab1}. Therefore, the complexified bilinear form $\mathcal{B}$ can be defined as
\beq\nonumber
        \mathcal{B}(\upxi, \upgamma_{{m_1 \ldots m_k}}\upxi) 
&=&  B(\upxi_{R}, \upgamma_{{m_1 \ldots m_k}} \upxi_{R}) 
-  B(\upxi_{I}, \upgamma_{{m_1 \ldots m_k}} \upxi_{I}) \\ &&\qquad +\, i \left( B(\upxi_{R}, \upgamma_{{m_1 \ldots m_k}} \upxi_{I})\right.\left. + B(\upxi_{I}, \upgamma_{{m_1 \ldots m_k}} \upxi_{R}). \right)
\eeq
\noindent The bilinear covariants can be now extended from the standard Majorana spinor field $\upxi \in \Gamma(\M,\mathcal{S})$ to sections of the $\Gamma(\M,\mathcal{S}_{\mathbb{C}})$, by setting the bilinear covariants now
as \begin{equation}
    \check{\mathcal{E}}^{(k)} = \frac{1}{k!}\mathcal{B}(\upxi, \upgamma_{{m_1 \ldots m_k}}\upxi)e^{m_1 \ldots m_k}. 
\end{equation}
\noindent Hence, since both terms in the real part and both terms in the imaginary part of the complex bilinear form can cancel each other,  the generalized bilinear covariants \eqref{nonvanishingbilinears} -- \eqref{nonvanishingbilinears2} can attain either non-vanishing or null values. Therefore $32$ new classes of spinor fields can be listed, according to the values of their generalized bilinear covariants. We display all the possibilities and analyze them below.

\begin{itemize}
    \item[1.] Five classes of spinor fields with one non-null bilinear covariant.
\begin{subequations}
\begin{align}
    &&&&\check{\mathcal{E}}^{(0)} \neq 0, &&&&\check{\mathcal{E}}^{(1)} = 0, &&&&\check{\mathcal{E}}^{(4)} = 0, &&&&\check{\mathcal{E}}^{(5)} = 0, &&&&\check{\mathcal{E}}^{(8)} = 0,\label{46a}\\
    &&&&\check{\mathcal{E}}^{(0)} = 0, &&&&\check{\mathcal{E}}^{(1)} \neq 0, &&&&\check{\mathcal{E}}^{(4)} = 0, &&&&\check{\mathcal{E}}^{(5)} = 0, &&&&\check{\mathcal{E}}^{(8)} = 0,\label{46b} \\
    &&&&\check{\mathcal{E}}^{(0)} = 0, &&&&\check{\mathcal{E}}^{(1)} = 0, &&&&\check{\mathcal{E}}^{(4)} \neq 0, &&&&\check{\mathcal{E}}^{(5)} = 0, &&&&\check{\mathcal{E}}^{(8)} = 0, \\
    &&&&\check{\mathcal{E}}^{(0)} = 0, &&&&\check{\mathcal{E}}^{(1)} = 0, &&&&\check{\mathcal{E}}^{(4)} = 0, &&&&\check{\mathcal{E}}^{(5)} \neq 0, &&&&\check{\mathcal{E}}^{(8)} = 0, \\
    &&&&\check{\mathcal{E}}^{(0)} = 0, &&&&\check{\mathcal{E}}^{(1)} = 0, &&&&\check{\mathcal{E}}^{(4)} = 0, &&&&\check{\mathcal{E}}^{(5)} = 0, &&&&\check{\mathcal{E}}^{(8)} \neq 0.
\end{align}
\end{subequations}
\end{itemize}

\begin{itemize}
    \item[2.] Ten classes of spinor fields, each one containing two non-null bilinear covariants:
\begin{subequations}
\begin{align}
    &&&&\check{\mathcal{E}}^{(0)} \neq 0, &&&&\check{\mathcal{E}}^{(1)} \neq 0, &&&&\check{\mathcal{E}}^{(4)} = 0, &&&&\check{\mathcal{E}}^{(5)} = 0, &&&&\check{\mathcal{E}}^{(8)} = 0, \\
    &&&&\check{\mathcal{E}}^{(0)} \neq 0, &&&&\check{\mathcal{E}}^{(1)} = 0, &&&&\check{\mathcal{E}}^{(4)} \neq 0, &&&&\check{\mathcal{E}}^{(5)} = 0, &&&&\check{\mathcal{E}}^{(8)} = 0, \\
    &&&&\check{\mathcal{E}}^{(0)} \neq 0, &&&&\check{\mathcal{E}}^{(1)} = 0, &&&&\check{\mathcal{E}}^{(4)} = 0, &&&&\check{\mathcal{E}}^{(5)} \neq 0, &&&&\check{\mathcal{E}}^{(8)} = 0, \\
    &&&&\check{\mathcal{E}}^{(0)} \neq 0, &&&&\check{\mathcal{E}}^{(1)} = 0, &&&&\check{\mathcal{E}}^{(4)} = 0, &&&&\check{\mathcal{E}}^{(5)} = 0, &&&&\check{\mathcal{E}}^{(8)} \neq 0, \\
    &&&&\check{\mathcal{E}}^{(0)} = 0, &&&&\check{\mathcal{E}}^{(1)} \neq 0, &&&&\check{\mathcal{E}}^{(4)} \neq 0, &&&&\check{\mathcal{E}}^{(5)} = 0, &&&&\check{\mathcal{E}}^{(8)} = 0, \\
    &&&&\check{\mathcal{E}}^{(0)} = 0, &&&&\check{\mathcal{E}}^{(1)} \neq 0, &&&&\check{\mathcal{E}}^{(4)} = 0, &&&&\check{\mathcal{E}}^{(5)} \neq 0, &&&&\check{\mathcal{E}}^{(8)} = 0, \\
    &&&&\check{\mathcal{E}}^{(0)} = 0, &&&&\check{\mathcal{E}}^{(1)} \neq 0, &&&&\check{\mathcal{E}}^{(4)} = 0, &&&&\check{\mathcal{E}}^{(5)} = 0, &&&&\check{\mathcal{E}}^{(8)} \neq 0, \\
    &&&&\check{\mathcal{E}}^{(0)} = 0, &&&&\check{\mathcal{E}}^{(1)} = 0, &&&&\check{\mathcal{E}}^{(4)} \neq 0, &&&&\check{\mathcal{E}}^{(5)} \neq 0, &&&&\check{\mathcal{E}}^{(8)} = 0, \\
    &&&&\check{\mathcal{E}}^{(0)} = 0, &&&&\check{\mathcal{E}}^{(1)} = 0, &&&&\check{\mathcal{E}}^{(4)} = 0, &&&&\check{\mathcal{E}}^{(5)} \neq 0, &&&&\check{\mathcal{E}}^{(8)} \neq 0,\\
    &&&&\check{\mathcal{E}}^{(0)} = 0, &&&&\check{\mathcal{E}}^{(1)} = 0, &&&&\check{\mathcal{E}}^{(4)} \neq 0, &&&&\check{\mathcal{E}}^{(5)} = 0, &&&&\check{\mathcal{E}}^{(8)} \neq 0.
\end{align}
\end{subequations}
\end{itemize}
\begin{itemize}
    \item[3.] Ten classes of spinor fields with three non-vanishing bilinear covariants:
 \begin{subequations}
\begin{align}
    &&&&\check{\mathcal{E}}^{(0)} \neq 0, &&&&\check{\mathcal{E}}^{(1)} \neq 0, &&&&\check{\mathcal{E}}^{(4)} \neq 0, &&&&\check{\mathcal{E}}^{(5)} = 0, &&&&\check{\mathcal{E}}^{(8)} = 0, \\
    &&&&\check{\mathcal{E}}^{(0)} \neq 0, &&&&\check{\mathcal{E}}^{(1)} \neq 0, &&&&\check{\mathcal{E}}^{(4)} = 0, &&&&\check{\mathcal{E}}^{(5)} \neq 0, &&&&\check{\mathcal{E}}^{(8)} = 0, \\
    &&&&\check{\mathcal{E}}^{(0)} \neq 0, &&&&\check{\mathcal{E}}^{(1)} \neq 0, &&&&\check{\mathcal{E}}^{(4)} = 0, &&&&\check{\mathcal{E}}^{(5)} = 0, &&&&\check{\mathcal{E}}^{(8)} \neq 0, \\
    &&&&\check{\mathcal{E}}^{(0)} \neq 0, &&&&\check{\mathcal{E}}^{(1)} = 0, &&&&\check{\mathcal{E}}^{(4)} \neq 0, &&&&\check{\mathcal{E}}^{(5)} \neq 0, &&&&\check{\mathcal{E}}^{(8)} = 0, \\
    &&&&\check{\mathcal{E}}^{(0)} \neq 0, &&&&\check{\mathcal{E}}^{(1)} = 0, &&&&\check{\mathcal{E}}^{(4)} \neq 0, &&&&\check{\mathcal{E}}^{(5)} = 0, &&&&\check{\mathcal{E}}^{(8)} \neq 0, \\
    &&&&\check{\mathcal{E}}^{(0)} \neq 0, &&&&\check{\mathcal{E}}^{(1)} = 0, &&&&\check{\mathcal{E}}^{(4)} = 0, &&&&\check{\mathcal{E}}^{(5)} \neq 0, &&&&\check{\mathcal{E}}^{(8)} \neq 0, \\
    &&&&\check{\mathcal{E}}^{(0)} = 0, &&&&\check{\mathcal{E}}^{(1)} \neq 0, &&&&\check{\mathcal{E}}^{(4)} \neq 0, &&&&\check{\mathcal{E}}^{(5)} \neq 0, &&&&\check{\mathcal{E}}^{(8)} = 0, \\
    &&&&\check{\mathcal{E}}^{(0)} = 0, &&&&\check{\mathcal{E}}^{(1)} \neq 0, &&&&\check{\mathcal{E}}^{(4)} \neq 0, &&&&\check{\mathcal{E}}^{(5)} = 0, &&&&\check{\mathcal{E}}^{(8)} \neq 0, \\
    &&&&\check{\mathcal{E}}^{(0)} = 0, &&&&\check{\mathcal{E}}^{(1)} \neq 0, &&&&\check{\mathcal{E}}^{(4)} = 0, &&&&\check{\mathcal{E}}^{(5)} \neq 0, &&&&\check{\mathcal{E}}^{(8)} \neq 0, \\
    &&&&\check{\mathcal{E}}^{(0)} = 0, &&&&\check{\mathcal{E}}^{(1)} = 0, &&&&\check{\mathcal{E}}^{(4)} \neq 0, &&&&\check{\mathcal{E}}^{(5)} \neq 0, &&&&\check{\mathcal{E}}^{(8)} \neq 0.
\end{align}
\end{subequations}
\end{itemize}
\begin{itemize}
    \item[4.] Five classes of spinor fields with four non-null bilinear covariants:
\begin{subequations}
\begin{align}
    &&&&\check{\mathcal{E}}^{(0)} \neq 0, &&&&\check{\mathcal{E}}^{(1)} \neq 0, &&&&\check{\mathcal{E}}^{(4)} \neq 0, &&&&\check{\mathcal{E}}^{(5)} \neq 0, &&&&\check{\mathcal{E}}^{(8)} = 0, \\
    &&&&\check{\mathcal{E}}^{(0)} \neq 0, &&&&\check{\mathcal{E}}^{(1)} \neq 0, &&&&\check{\mathcal{E}}^{(4)} \neq 0, &&&&\check{\mathcal{E}}^{(5)} = 0, &&&&\check{\mathcal{E}}^{(8)} \neq 0, \\
    &&&&\check{\mathcal{E}}^{(0)} \neq 0, &&&&\check{\mathcal{E}}^{(1)} \neq 0, &&&&\check{\mathcal{E}}^{(4)} = 0, &&&&\check{\mathcal{E}}^{(5)} \neq 0, &&&&\check{\mathcal{E}}^{(8)} \neq 0, \\
    &&&&\check{\mathcal{E}}^{(0)} \neq 0, &&&&\check{\mathcal{E}}^{(1)} = 0, &&&&\check{\mathcal{E}}^{(4)} \neq 0, &&&&\check{\mathcal{E}}^{(5)} \neq 0, &&&&\check{\mathcal{E}}^{(8)} \neq 0, \\
    &&&&\check{\mathcal{E}}^{(0)} = 0, &&&&\check{\mathcal{E}}^{(1)} \neq 0, &&&&\check{\mathcal{E}}^{(4)} \neq 0, &&&&\check{\mathcal{E}}^{(5)} \neq 0, &&&&\check{\mathcal{E}}^{(8)} \neq 0.
\end{align}
\end{subequations}
\end{itemize}

\begin{itemize}
    \item[5.] A single class of spinor fields with all five non-vanishing bilinear covariants. 
    \begin{align}
&&&&\check{\mathcal{E}}^{(0)} \neq 0, &&&&\check{\mathcal{E}}^{(1)} \neq 0, &&&&\check{\mathcal{E}}^{(4)} \neq 0, &&&&\check{\mathcal{E}}^{(5)} \neq 0, &&&&\check{\mathcal{E}}^{(8)} \neq 0.\label{50}
\end{align}
\end{itemize}

\begin{itemize}
    \item[6.] A single class consisting of null bilinear covariants (trivial class):
        \begin{align}
&&&&\check{\mathcal{E}}^{(0)} = 0, &&&&\check{\mathcal{E}}^{(1)} = 0, &&&&\check{\mathcal{E}}^{(4)} = 0, &&&&\check{\mathcal{E}}^{(5)} = 0, &&&&\check{\mathcal{E}}^{(8)} = 0. \label{trivialclass}
\end{align}
\end{itemize}

Ref. \cite{Bon14} proposed new classes of spinor fields on the $S^7$ component of the AdS$_4\times S^7$ compactification scheme, also hinging upon geometric Fierz identities, whose structure obstructs the number of non-vanishing bilinear covariants on the $S^7$ sphere. Nevertheless,  three non-trivial new emergent sets of fermionic fields were implemented on $S^7$. 
Refs. \cite{Bon14,Yanes:2018krn} also pointed to those new classes of spinor fields and to the latest obtained fermionic solutions of first-order equations of motion, playing a significant role in new trends in AdS/CFT and supergravity. For the AdS$_3\times \M_8$ compactification here studied, Eq. (\ref{50}) has been already reported in Ref. \cite{bab2}. The bilinear form  \eqref{bc} 
makes it possible to implement 32 additional spinor field classes, ulterior to the one in Ref. \cite{bab2}. The spinor bilinear forms constituting the Fierz aggregate were employed in Ref. \cite{Bonora:2017oyb} to successfully provide a second-quantized spinor field classification on 4-dimensional Lorentzian manifolds. In the perturbative approach to quantum field theory, the propagators make possible the computation of correlators through the Wick contraction theorem. In any theory encoding interactions among their constituents, propagators are additionally fundamental tools to evaluate correlators. The classification of quantum spinor fields according to their bilinear covariants provided the subsequent classification of propagators, constructed upon regular and singular spinor field classes (\ref{tipo1}) -- (\ref{tipo6}) \cite{Bonora:2017oyb}. 
 A quantum reconstruction algorithm was also established in Ref. \cite{Bonora:2017oyb}, with the Feynman propagator extended for all sets of quantum spinor fields. This idea can be implemented to also establish the second-quantized scheme in the spinor field classification arising in the context of flux compactification AdS$_3\times \M_8$, heretofore discussed.

\section{Conclusions} \label{sec7concl}
Using the K\"ahler--Atiyah bundle and the Clifford bundle tools, generalized geometric Fierz identities were derived, based on the admissible pairings on the bundle of real spinors. Building upon Lounesto's spinor field classification in Minkowski spacetime, we extended the spinor field classification to the warped $\mathcal{N} = 1$ supersymmetric compactification  AdS$_3 \times \M_8$. Algebraic and differential obstructions  cause some of the bilinear covariants to vanish, leading to the identification of new classes of spinor fields. There exists a distinction between Lounesto's classification and the conventional global classification of spinor fields. The global classification offers a unified framework for spinor fields defined on a manifold, Lounesto's classification operates on a pointwise basis, not providing a globally classification of spinor fields. This limitation arises due to the existence of locations on the manifold where quantities relevant to Lounesto's classification may transition abruptly from zero to nonzero values. Such discrepancies are evident in the equations presenting Lounesto's classification and the new 32 classes of spinor fields, where quantities compared to zero depend on the specific point on the manifold in geometric applications.\color{black} The supersymmetry conditions imply that one can derive the general form of solutions from information extracted from the constrained generalized Killing spinor equations in $\M_8$  \cite{Kim:2005ez}.   Bilinear covariants and generalized Fierz identities arise from the 
constrained generalized Killing spinor equations. For lower-dimensional systems with a lower number of  supersymmetries, in some cases the classification of all supersymmetric solutions is possible, with several important classes of new solutions reported in Ref. \cite{Kim:2005ez}. The discovery of regular and singular spinor field classes in Lounesto's classification paved new proposals for the construction of  fermionic fields in several physical backgrounds. Further classes of spinor field have been emulated in the AdS$_5\times S^5$ and AdS$_4\times S^7$ flux compactifications, in Refs. \cite{Bon14,brito,Lopes:2018cvu, Yanes:2018krn,Meert:2018qzk}, providing concrete new directions to engender fermionic solutions in supergravity, string theory, and gauge/gravity dualities. Finding 32 new classes of spinor fields in the AdS$_3\times \M_8$ compactification extends the previous results and can accommodate new supersymmetric fermionic solutions.

\textcolor{black}{The AdS/CFT duality is currently one of the best tools for exploring and studying
quantum gravity.  The only way to define quantum gravity from a non-perturbative aspect is through CFT, allowing precise  calculations. Hence, several questions regarding quantum gravity can be answered using CFTs, if one knows how to implement the right calculations, perturbatively or even non-perturbatively. 
Employing the methods underlying string theory and AdS/CFT, one can tackle complex problems in condensed matter physics in innovative ways, also offering new phenomenological perspectives. These approaches allow us to transform intricate tasks into more manageable ones, which can therefore be solved in a more straightforward way,  shedding new light on their relevance to, e.g., condensed matter physics and providing a constructive dialogue between it and the dual gravity framework.
Even before the seminal work \cite{Herzog:2007ij}, string theory and condensed matter had already met each other on the quantum criticality/conformal field theory duality (AdS/CMT). 
Since AdS/CFT and its version AdS/QCD have shed new light on QCD and the quark-gluon plasma, with new phenomenological aspects provided by the holographic duality, AdS/CMT is one of the propitious scenarios for exploring the holographic dictionary, including the possibility of exploring new phenomena involving superconductors and 2-dimensional topological insulators, holographic quantum critical phases, strange metals, and non-Fermi liquids, as the graphene 
(see, e.g.,  \cite{Zaanen:2015oix,Nastase:2017cxp} and references therein). Due to the infinite-dimensional symmetry of two-dimensional CFTs, AdS$_3$/CFT$_2$ is one of
the few locations where one can perform explicit calculations, in particular involving the spin-chain and the Bethe ansatz  \cite{OhlssonSax:2011ms}. 
We want to shed new light on the fermionic sector of fluid/gravity duality, in the context of the AdS$_3\times \M_8$ compactification. As this endeavour is a comprehensive one in itself, we have paved in this work the formal aspects for further developments.}

\section*{Acknowledgements}
DFG thanks to the Coordination for the Improvement of Higher Education Personnel (CAPES - Brazil, Grant No. 001) and the Center for Mathematics of the Federal University of ABC. RdR~thanks to The S\~ao Paulo Research Foundation -- FAPESP
(Grants No. 2021/01089-1 and No. 2024/05676-7) and to the National Council for Scientific and Technological Development -- CNPq  (Grants No. 303742/2023-2 and No. 401567/2023-0), for partial financial support; to Prof. Jorge Noronha and the Illinois Center for Advanced Studies of the Universe, University of Illinois at Urbana-Champaign, for the hospitality; and to Prof. Roberto Casadio and DIFA --  Universit\`a di Bologna, for the hospitality.




\end{document}